\documentclass[aps,prb,amsmath,amssymb,lengthcheck,showpacs,superscriptaddress]{revtex4-1}
\usepackage{color}
\usepackage{graphicx}
\begin{document}
\title{Polymer translocation through nano-pores in vibrating thin membranes}
\author{Timoth\'ee Menais}
\author{Stefano Mossa}
\author{Arnaud Buhot}
\email{arnaud.buhot@cea.fr}
\affiliation{Univ. Grenoble Alpes, INAC-SYMMES, F-38000 Grenoble, France}
\affiliation{CNRS, INAC-SYMMES, F-38000 Grenoble, France}
\affiliation{CEA, INAC-SYMMES, F-38000 Grenoble, France}
\date{\today}
\begin{abstract}
Polymer translocation is a promising strategy for the next-generation DNA sequencing technologies. The use of biological and synthetic nano-pores, however, still suffers from serious drawbacks. In particular, the width of the membrane layer can accommodate several bases at the same time, making difficult accurate sequencing applications. More recently, the use of graphene membranes has paved the way to new sequencing capabilities, with the possibility to measure transverse currents, among other advances. The reduced thickness of these new membranes poses new questions on the effect of deformability and vibrations of the membrane on the translocation process, two features which are not taken into account in the well-established theoretical frameworks. Here, we make a first step forward in this direction. We report numerical simulation work on a model system simple enough to allow gathering significant insight on the effect of these features on the average translocation time, with appropriate statistical significance. We have found that the interplay between thermal fluctuations and the deformability properties of the nano-pore play a crucial role in determining the process. We conclude by discussing new directions for further work. 
\end{abstract}
\maketitle
\section{Introduction}
Translocation processes of biomolecules (in particular, DNA) through nano-pores involve interesting experimental and theoretical issues, at the center of an intense activity. The translocation process consists in a bio-molecule crossing a membrane through a hole of nano-metric size, called nano-pore. This process can either be natural (unbiased) or forced due to electrophoretic interactions for example. Experimentally, the first successfully forced DNA translocation through a biological nanopore was obtained in 1996 by Kasianowicz {\em et al.}~\cite{Kasianowicz}. Since then, translocation phenomena have been used to characterize single objects like single-stranded or double-stranded DNA, proteins, or even cells.~\cite{Dekker, Keyser, Meller,Movileanu, Palyulin, Wanunu, Liu} High impact potential applications in biotechnologies and medical diagnostics are expected, with a clear emphasis on quicker and cheaper methods for DNA sequencing.~\cite{Branton, Clarke, Merchant, Schneider2}

Current limitations to the use of DNA translocation as a sequencing tool are both temporal and spatial. First, in most experiments a base spends about 1 $\mu$s within the pore, while currents measurements resolution times would require a slowing down of the process leading to an occupation time of around 1 ms.~\cite{Branton} Several options have been explored to reduce the translocation speed~\cite{Carson}, including modifying the electrolyte solution by adding large amounts of Lithium salts~\cite{Kowalczyk} or glycerol,~\cite{Fologea} or by reducing the nano-pore size.~\cite{Mirsaidov, Schneider} Second, in the case of common biological and artificial nano-pores, several nucleotides are simultaneously present within the nanopore during the translocation, hampering the possibility of single-base sequence resolution.~\cite{Schneider}

In the last decade, outstanding experiments have involved translocations through nanopores carved in mono-atomic graphene sheets~\cite{Garaj, Merchant, Schneider}. These new membranes may provide a solution for the thickness issue, since their width is substantially smaller than a nucleotide. The possibility to measure transverse currents may also improve the detection capabilities~\cite{Nelson, Prasongkit, Saha, Traversi}. The graphene sheet, however, is flexible and may suffer from vibrations due to both thermal and elastic fluctuations. More recently, an additional range of membranes based on DNA origamis~\cite{Rothemund} has appeared. These soft membranes are constituted by two-dimensional sheets, formed by the hybridization of small strands (staples) of DNA with a long circular scaffold strand generally coming from a virus genome. By selecting small staple strands, the membrane may contain nano-pores of tuned size~\cite{Bell, Hernandez, Hernandez2, Wei}, whose flexibility and deformability may affect the translocation process. 

Interestingly, the above extremely tiny materials challenge a crucial assumption which is commonly made in most theoretical approaches, where the membrane and the nano-pore are considered {\em immobile}~\cite{Palyulin}. This constraint is acceptable in former experiments with thick membranes, where the impact of vibrations and flexibility on the translocation process can indeed be neglected. In the case of thin and/or soft membranes, however, these effects are expected  to strongly influence the translocation time of biomolecules. Similarly, although important numerical simulation work has been published on the statistical physics of the translocation process (see the reviews of Refs.~\cite{Milchev, Palyulin} and references therein), relatively limited attention has been paid to the case of a vibrating membrane. Following experimental results, numerical simulations have been devoted to understand DNA translocation through nano-pores carved in graphene membranes. Most of them, unfortunately, are based on {\em ab-initio} calculation, mainly dealing  with the problem of bases discrimination~\cite{Nelson, Prasongkit, Saha}. On the other side, classical Molecular Dynamics (MD) simulation based on all-atoms descriptions are limited in the number of translocation instances generated, and a general picture could not be deduced simply due to a lack of appropriate statistical significance.~\cite{Li, Qiu, Wells, Zhang}

Here, we report classical MD simulation work of a simple coarse-grained model for both the translocating polymer and the membrane. These incorporate all the necessary ingredients to clarify the effect of the kinetics of the membrane on the translocation time probability distributions. The model polymer we have considered is flexible and presents a base structure mimicking single-stranded DNA at the coarse-grained level. The membrane is characterized by the hexagonal lattice structure of graphene-like sheets, with dynamical interaction sites tethered to the reference lattice by harmonic wells. The models are therefore simple enough to allow for a statistically significant sampling of translocation processes, and, therefore, for general conclusions on the effect of membrane deformations and thermal vibrations.
\begin{table}[b]
\begin{tabular*}{0.49\textwidth}{llll}
\hline
\hline
Observable&Dimensions& LJ units& SI units\\
\hline
\hline
Length & $L$ & $a$ & $0.3$ nm \\
Mass & $M$ & $m$ & $1.6 \times 10^{-25}$ kg \\
Energy & $E$ & $\epsilon$ & $2.74 \times 10^{-21}$~J\\
Temperature ($T_p$) & $T$ & $3/2\epsilon$ & $300$~K\\
Force & $E/L$ & $\epsilon/a$ & $9.1$~pN\\
Time & $\sqrt{ML^2/E}$ & $\sqrt{m a^2/\epsilon}$& $2.3$~ps\\
Friction & $ L^3/E T$ & $\xi$ & $7.6 \times 10^{-9} m^{3} (JK)^{-1}$\\
\hline
\hline
\end{tabular*}
\caption{
Main observables (with the appropriate dimensions) in Lennard Jones units, together with typical associated 
values in SI units for the mimicked real system. Note that $k_B=1$ in LJ units.
}
\label{tbl:example}
\end{table}
\section{Results}
\subsection{Model and simulation details}
\subsubsection{The structured polymer} 
\begin{figure}[t]
\centering
\includegraphics[width=0.48\textwidth]{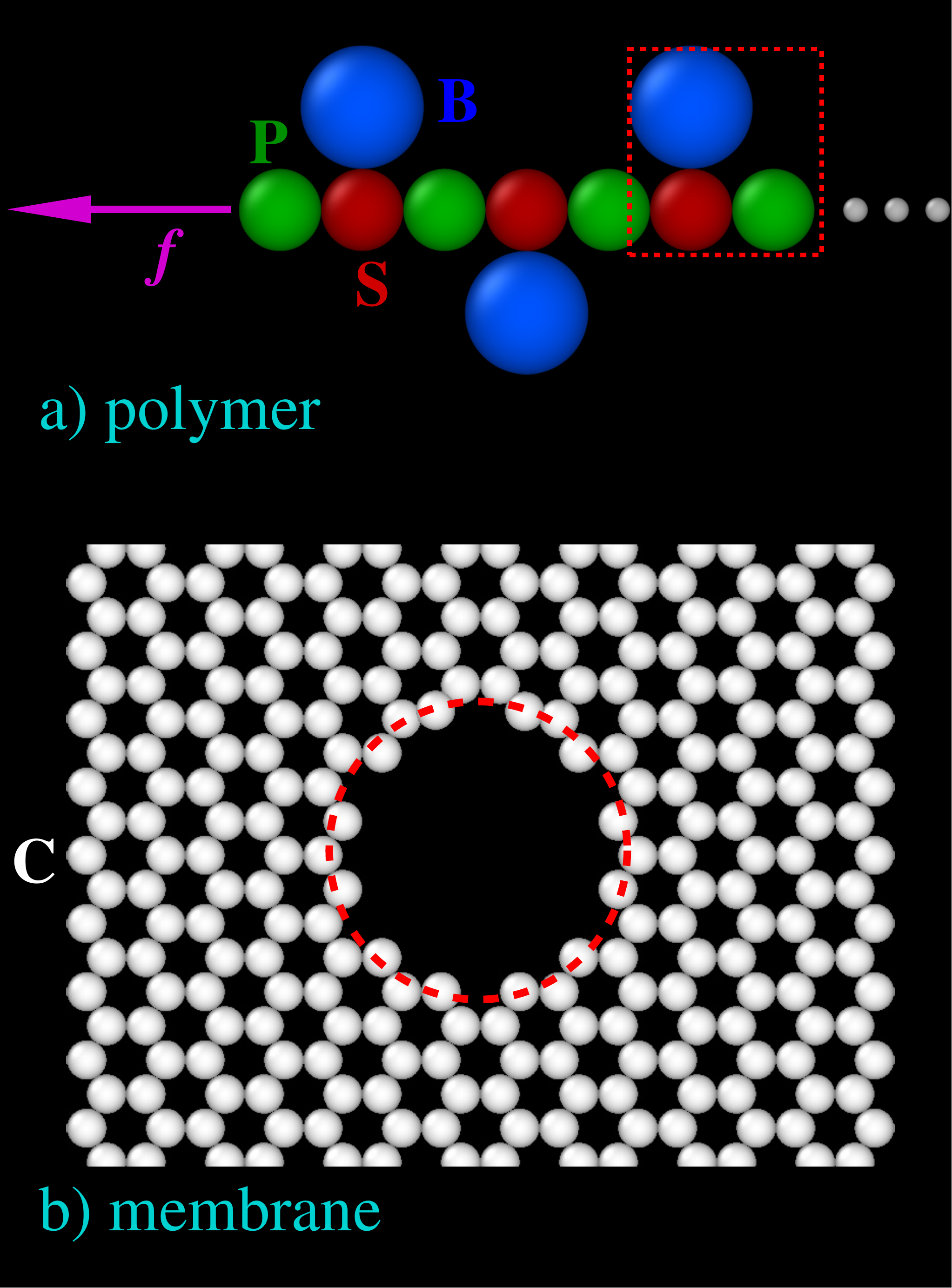}
\caption{
Sketches of {\em a)} the coarse-grained models developed for the structured translocating polymer (mimicking a DNA single-strand) and {\em b)} the nano-pore carved in the graphene-like membrane. The P, S and B connected beads (dashed red square) constitute the fundamental building block which is replicated to form the polymer. The constant pulling force $f$ (our control parameter) acts in the $z$-direction along the pore axis on the P-bead added at the left-most extremity of the chain. The nano-pore (dashed red circle) is carved in the membrane by removing the C-beads placed at distances less than $R_{s,l}$ from the center (origin).
}
\label{fig:model}
\end{figure}

We have devised a minimalistic bead-spring polymer model for single-stranded DNA, only including sterical repulsion and binding of the monomers (see Fig.~\ref{fig:model}a)). Inspired by the sensibly more elaborated description of Ref.~\cite{Linak}, we have introduced three types of beads: $n_P$ phosphate-like (P) and $n_S$ sugar-like (S) units are alternated to form the polymer backbone, while $n_B$ lateral base-like (B) units are grafted to the S-beads (Fig.~\ref{fig:model}a)). No distinction is made at this level between the four bases. Obviously $n_S=n_B=n$, while one additional P-bead is present at the beginning of the chain ($n_P=n+1$). This unit is also subjected to the application of the pulling force, as we will see below. The total number of beads in the polymer is therefore $N=3n+1$. While we have explored a range of $N$, in the following we will discuss results for $N=49$ ($n=16$).

Any pairs of monomers in the chain are subjected to steric interactions, modeled by a truncated and shifted Lennard Jones (LJ) potential,
\begin{equation}
U_{LJ}(r_{ij})= 4\epsilon \left[\left(\frac{\sigma_{ij}}{r_{ij}}\right)^{12}-\left(\frac{\sigma_{ij}}{r_{ij}
}\right)^{6}\right] + \epsilon,
\label{eqn:lj}
\end{equation}
for $r_{ij}\le 2^{1/6} \sigma_{ij}$, and $0$ otherwise. Here, the energy scale $\epsilon=1$, $r_{ij}$ is the distance between beads $i$ and $j$, and $\sigma_{ij} = (\sigma_i + \sigma_j)/2$ with $\sigma_i$ the radius of bead $i$. $U_{\text {LJ}}$ is a purely repulsive potential, which only prevents the beads to overlap at exceedingly short distances. We have chosen $\sigma_S= \sigma_P=a=0.3$~nm and $\sigma_B = 1.5\; a$ ($0.45$~nm). S-P and S-B pairs which are adjacent and, therefore, connected, also exert a mutual force consistent with the potential,
\begin{equation}
U_B(r_{ij})= -15\epsilon \left(\frac{R_{ij}}{\sigma_{ij}}\right)^2\text{ } \ln\left[1-\left(\frac{r_{ij}}{R_
{ij}}\right)^2\right],
\end{equation}
with the value of the bond length $R_{ij} = 1.5\;\sigma_{ij}$. This term is attractive and extends to $R_{ij}$, the maximum elongation of the bond. The sum $U_{LJ}+U_{B}$ is the FENE potential~\cite{Kremer}. Finally, all beads have the same mass $m=1$ in LJ units. In Table~\ref{tbl:example} we report the LJ rescaled values of all important quantities, together with the associated typical values in SI units. 

The translocation is conducted under the influence of a constant time-independent force $\mathbf{F} = F \hat{z}$ acting on the first P-bead of the chain (see Fig.~\ref{fig:model}a)). In the following we will consider the rescaled force $f=F a / k_B T_p$ as our control parameter, with $T_p=3/2$ the polymer temperature (see below for details). This procedure mimics set-ups considered in recent experimental work and based on optical or magnetic tweezers~\cite{Keyser1, Keyser2, Peng, Trepagnier, vanDorp}.   

We have verified that the behavior of our polymer model is consistent with the main exact predictions for the case of the free chain~\cite{Grosberg, Doi}. In particular, we have confirmed that the diffusion constant scales as the inverse of the polymer length ($D\propto N^{-1}$), due to the lack of hydrodynamical interactions. Friction of the polymer also scales with the length, and the fluctuation-dissipation theorem relating the two variables is also fulfilled~\cite{Grosberg, Doi}. Finally, the polymer gyration radius scales as function of the number of nucleotides, with a Flory exponent $\nu \simeq 0.69$, to be compared with the expected value $\nu = 0.588$. The non-trivial structure of the polymer, with the presence of side chains, and finite size effects explain this discrepancy and apparent larger exponent~\cite{Grosberg}. 

We conclude by noting that with this minimalistic model, we cannot consider the effect of mechanisms like hydrogen bonding, base stacking or backbone bending. As a consequence, we are not able to tackle issues such as, for instance, stiffness and helicoidal structure building of the DNA double helix~\cite{Linak}. However, since we focus on short single-stranded DNA translocation without secondary structures formation like hairpins, these degrees of freedom are expected to be irrelevant in the present context. Their implementation would have furthermore implied additional heavy computational cost, at the expense of an accurate translocation sampling. For similar reasons, no explicit solvent is considered and hydrodynamic interactions are neglected. 
\subsubsection{The membrane} 
Graphene is a 2-dim crystal formed by aromatic carbon cycles arranged on a honeycomb lattice. A sketch of the membrane with the carved nano-pore we have considered is shown in Fig.~\ref{fig:model}b). Here the carbon atoms (C-beads) are represented as white spheres, and the membrane is comprised in the $x-y$ plane, with the normal $\hat{n}=\hat{z}$. In order to keep realistic size ratios of the graphene atoms to the (DNA) polymer beads, we have fixed the hexagonal lattice constant to a value $b=a/2=0.15$~nm. Periodic boundary conditions are applied in the $x-y$ plane only, while open boundaries are used in the $z-$direction, such that the box extent in this direction follows at each time step the translocation process. The membrane size is chosen large enough to avoid interaction of a fully stretched DNA strand with its own image in the $x-y$ plane, implying considering a large number of interaction sites for the membrane ($N_C=5488$ carbon atoms). The nano-pore trough which the translocation takes place is carved in the membrane (red dashed circle in Fig.~\ref{fig:model}b)), by erasing the $n_C=24$ atom carbons at distances less than the pore radius $R_s = 1.25\,a$ from the $(0,0,0)$ reference point. We have also considered the additional case of a larger radius $R_l \simeq 1.75\,a$ erasing $n_C = 54$ carbon atoms (static membrane, see below). Membrane beads interact among themselves (in the dynamical case, see below) via the LJ potential of Eq.~(\ref{eqn:lj}) with $\sigma=a/3=0.1$~nm, and with the polymer beads with $\sigma=a$. This last choice for the membrane-polymer interactions assures that the polymer cannot penetrate the membrane sneaking between adjacent membrane beads. The polymer is thus constrained inside the nano-pore during the entire translocation process.

In this study we have considered the two cases of i) {\em immobile} and ii) {\em kinetic} membrane. In the former reference case, the C-atoms pertaining to the membrane are not allowed to move. In the dynamical case, in contrast, the C-atoms at position $\mathbf{r}$ are tethered to the respective equilibrium positions $\mathbf{r}_o$  in the hexagonal lattice, by harmonic potentials of the form $U_{\text{harm}}=k/2|\mathbf{r}-\mathbf{r}_o|^2$, with a trap strength $k$ (elastic constant). They can therefore move undergoing small displacements from their equilibrium positions, following both coupling with the thermostat and direct interactions with the other beads. Obviously we expect a larger contribution to the process coming from the beads closer to the pore, in direct interaction with the polymer. A non-trivial effect, however, is also expected from membrane atoms further from the pore, which exchange momentum with the polymer monomers before translocation.
  
To precisely assess the effect of vibration and deformability on translocation times, a large range of trap strengths have been considered, from $k = 75$ up to $k = 10^4$ (LJ units), with the fixed membrane reference case corresponding to $k=\infty$. In order to clarify the role played by the natural frequency of oscillation of the membrane beads $\omega_C^2=k/m_C$, the mass of the membrane beads was varied from $m_C=10^{-1}$ to $10$ in LJ units (see Table~\ref{tbl:example}). We observed very minor effects on the translocation time due to mass variation. All the data presented below therefore correspond to $m_C=m=1$. 
\begin{figure*}[t]
\centering
\includegraphics[width=0.95\textwidth]{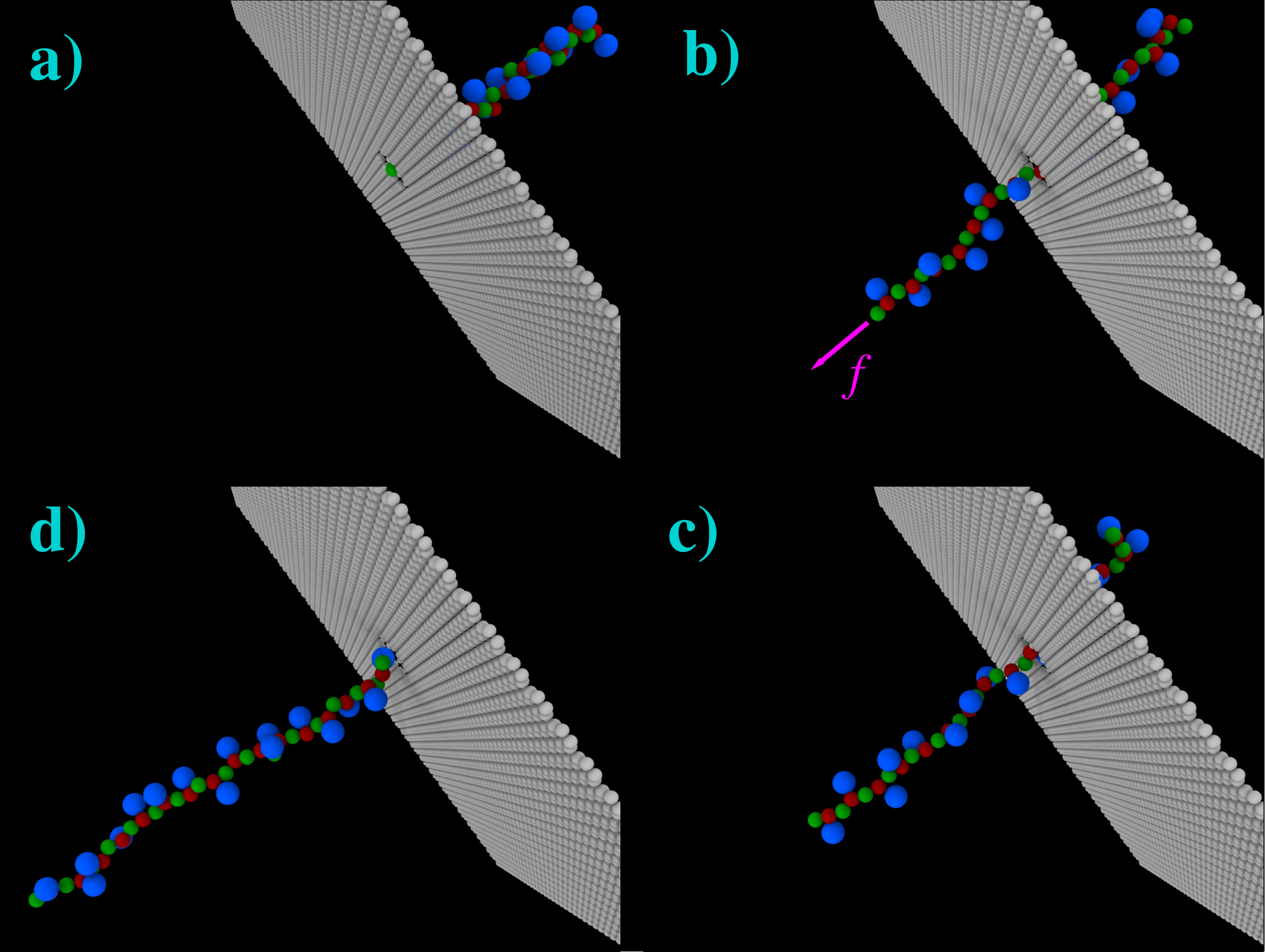}
\caption{
Snapshots at subsequent times (from a) to d)) of a typical polymer translocation instance. Snapshot a) follows the initialization step, performed by immobilizing the first monomer at the center of the nanopore, and letting the polymer explore the conformations space restricted by the presence of the membrane at the cis side ($z < 0$). At time $t=0$ a) the pulling force is switched on, and the translocation process starts. The translocation is considered as completed, d), when all the polymer beads are on the trans side of the membrane. The colors for the beads are those of Fig.~\ref{fig:model}.
}
\label{fig:snaps}
\end{figure*}
\subsubsection{Simulation details} 
The dynamics of each bead is determined by a Langevin equation of the form
\begin{equation}
m\frac{\partial^2 \mathbf{r}}{\partial t^2}=-\nabla U(\textbf{r})-\xi \frac{\partial \mathbf{r}}{\partial t}
+{\boldsymbol \eta},
\label{eq:langevin}
\end{equation}
with $\textbf{r}$ the position vector of the bead, $\xi=1$ the friction coefficient and $\boldsymbol{\eta}$ the noise, such that $\left<\eta_i (t) \eta_j (t') \right> = 2 \xi k_B T \delta_{ij}\delta(t-t')$. We have coupled separately the polymer and the membrane to Langevin thermostats, to be able to consider different temperatures. We have fixed $T_p = 3/2\,\epsilon$ for the polymer, while the temperature $T_m$ for the membrane beads has been varied in the range $0$ to $T_p$. The equations of motion where integrated numerically with a time step $\delta t = 5 \times 10^{-3}$ (about $26$~fs in SI units). Each translocation instance is initialized by immobilizing the first monomer at the center of the nanopore ($\mathbf{x}_0 = (0,0,0)$), followed by  a thermalization run where the conformations space is restricted by the presence of the membrane at the cis side ($z < 0$). At time $t=0$ the pulling force ${\bf F} = F {\hat z}$ is activated and the translocation starts. The process is considered as terminated when all the beads of the polymer are on the trans side ($z>0$). The time step corresponding to termination is the translocation time. At each pulling force $f$, we have performed $N_t = 10^3$ translocations, constituting the statistical ensemble for the subsequent analysis. 

We have investigated the effect of pulling forces $f= Fa/k_B T_P$ in the range from $0.2$ to $60$, corresponding to $F = 2$ to $600$~pN. In most cases at the lower forces, the external driving is not sufficient to pull in the nano-pore the polymer, which therefore starts to slide against the membrane, rolling away from the pore. These trajectories are obviously disregarded. In Fig.~\ref{fig:snaps} we show a series of snapshots of the system, dumped at increasing times (clock-wise from a) to d)) during a typical translocation process. All simulations have been performed by using the high-performance parallel computing Molecular Dynamics code LAMMPS~\cite{Plimpton}. In what follows we focus on the behavior of the average translocation time $\tau$ in the different considered conditions. The values of $\tau$ have been evaluated, together with the errors on the average ($\sigma^2/\sqrt{N_t}$ ), from successful translocation events for each state point. In all cases the probability distribution of the translocation times follows a first-time-passage diffusion-like distribution proposed by Ling and Ling~\cite{Ling}. We show a few examples of our data in Supplementary Fig. S1 online.
\begin{figure}[t]
\centering
\includegraphics[width=0.48\textwidth]{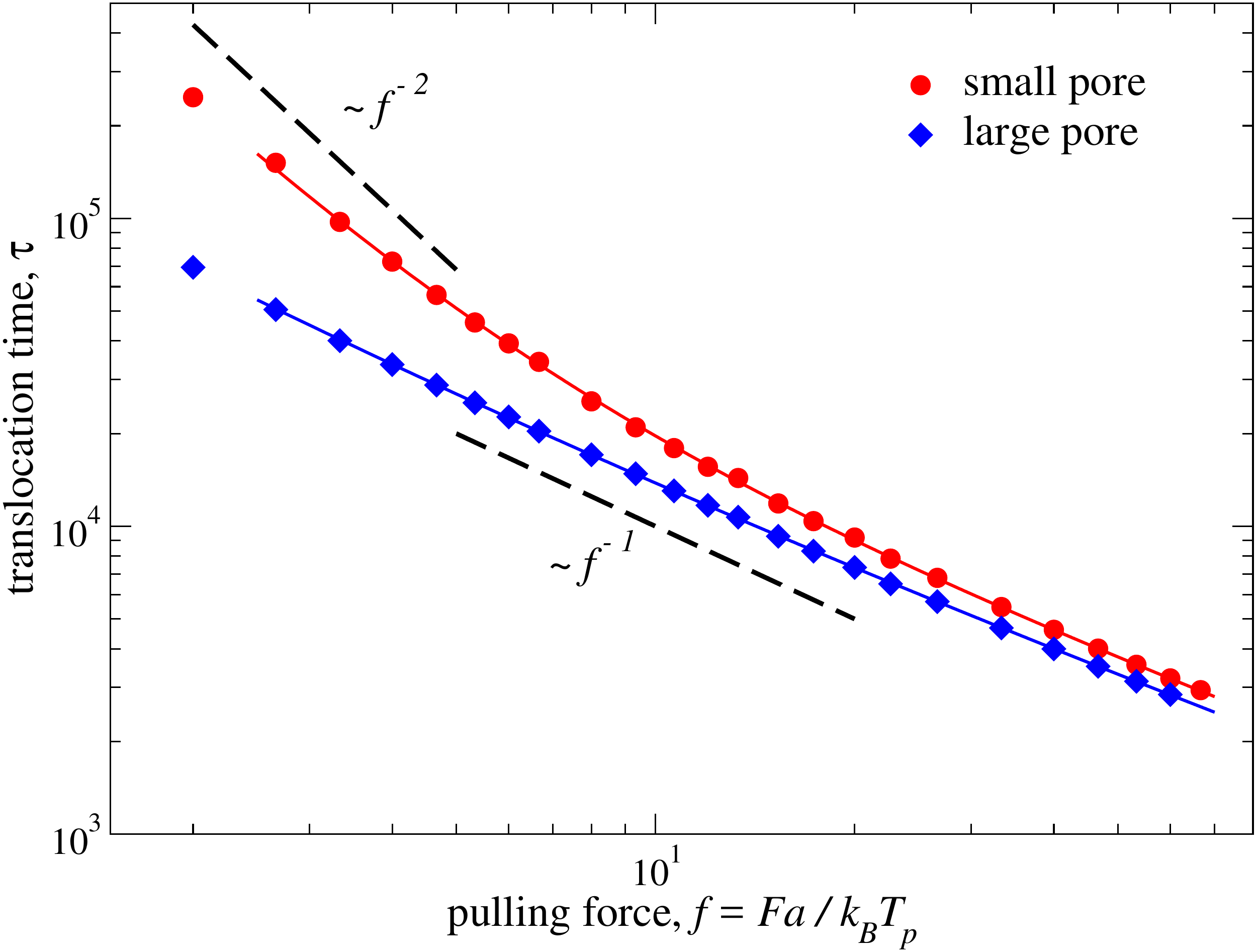}
\caption{
Average translocation times, $\tau$ with small ($R_s$) and large ($R_l$) nanopores for the structured polymer of size $N=49$, as a function of the rescaled pulling force $f$. Decreasing the size of the nano-pore slows down the translocation process at fixed $f$. At large $f$ modifications are limited, whereas at low forces the size of the pore strongly influences the translocation. The translocation time scales as about the inverse of the force $\propto f^{-1}$ (bottom dashed line) for the large pore in the entire range, with a slight upward bending at high-$f$. For the narrow pore this behavior is observed at high-$f$, while we find a cross-over to an apparent $\tau\propto f^{-2}$ in the low-force regime (top dashed line). 
}
\label{fig:small-large}
\end{figure}
\subsection{Static effect of the size of the nano-pore}
One of the most severe issues in sequencing applications is related to the speed of translocation. Indeed, current sensibility in measurement techniques is challenged by a process that is exceedingly fast in most experiments~\cite{Branton}. We start by investigating the effect of the pore size on $\tau$ in the static case, with the membrane kept fixed.  We have first evaluated the average translocation time though a nano-pore of radius $R_l$, which is large enough to fulfill the hypothesis of most theoretical approaches, i.e., a few beads may occupy the pore at the same time, and no additional friction is opposed by the pore due to polymer deformation or polymer-membrane interactions~\cite{Milchev}. Our results are shown in Fig.~\ref{fig:small-large} (diamonds) and do not reserve particular surprises. Clearly, $\tau(f)$ decreases with $f$ in the entire considered range, for an overall variation of almost two orders of magnitude. Also, the data generally conform to the expected behavior $\tau(f)\propto f^{-\gamma}$, with $\gamma=1$,~\cite{Meller, Palyulin} with a minor upward bending due to a slight modification of $\gamma\simeq0.9$ for $f>20$. More in details, the observed $1/f$-behavior for the present large nano-pore is principally due to the friction (or drag) force between the polymer and the solvent, implicitly included in the Langevin equation. The velocity of the polymer is therefore proportional to the exerted force, leading to an average translocation time inversely proportional to $f$. The observed slight decrease of the exponent $\gamma$ is instead generally attributed to the polymer bonds stretching at large values of the pulling force. This last observation can be easily rationalized, by observing that the linear response requirement is certainly violated at these large values of $f$.

In the case of the vibrating membrane, we expect that the instantaneous (effective) size of the pore, $R_{\text{eff}}(t)$, will vary during the translocation, following both thermal agitation and direct interaction with polymer and membrane beads. In the case of $R_{\text{eff}}(t)>R_l$, we obviously do not expect any variation compared to the above data. For $R_{\text{eff}}(t)<R_l$, however, the situation should change due to additional steric effects, summing up in an increased friction presented by the pore. We can be more precise on this point, by considering a smaller pore of radius $R_s$ with a fixed membrane. The size of the pore $2 R_s = 2.5\,a$ is now equivalent to the size of a backbone sugar and its adjacent base ($\sigma_S+\sigma_B$). This imposes some deformation of the polymer during the translocation and stronger polymer-membrane interactions. Our data are shown in Fig.~\ref{fig:small-large} (circles) and now exhibit a non-trivial behavior, which is substantially more complex than the above and illustrates the influence of those interactions. 

First, we find the expected overall increase of $\tau$ compared to the large pore case. This is simple to rationalize, by observing that in this case the polymer B-beads must tilt to cross the pore (see Fig.~\ref{fig:model}). As a consequence, the energy landscape associated to the translocation process is strongly modified compared to the large nanopore, with a substantial increase of the free energy barriers opposing the translocation. Also, the high-$f$ behavior is still consistent with $\propto f^{-\gamma}$ with $\gamma\simeq 1$, and the two sets of data are very close. This is also expected: at high pulling force translocation is actually dominated by this latter and, as a consequence, pore friction only plays a secondary role. More interestingly, at intermediate values of $f$ we find a clear cross-over to a slower behavior, with an apparent $f^{-2}$ dependence at low-$f$. The effect of the pore size is therefore larger at smaller pulling forces, amounting to a slow-down of more than a factor $4$ at $f=2$. 

An observation is in order at this point. The data points for $f\le4$ are also consistent with an exponential behavior $\propto \exp{(-\alpha f)}$ at low-$f$, which points to an activated translocation process. This finding is coherent with the presence of stalling periods in the polymer translocation, which clearly correlate with the translocation of the bases grafted to the polymer backbone. (We show an example of this point in Supplementary Fig.S3 online.) Indeed, due to the sizes of the backbone and base beads, whose sum is close to the pore diameter, translocation of the polymer usually requires deformation of the backbone or (tilt) motion of the base engaging the pore. This last process is on average slower than the former and therefore signals the passage of particular components of the polymer, clearly demonstrating the potential for DNA sequencing. Similar results have been recently observed also for proteins~\cite{Wilson}. The presence of knots or foldings of the polymer may also cause intermittent stalling periods on the translocation, as recently demonstrated in Refs.~\cite{Szymczak,Bonome}. In the Supplementary Fig. S2 online, we also show the single bead waiting time inside the pore as a function of the translocation coordinate (bead index), which points even more clearly toward the intermittent character of the observed dynamical process.  

The overall picture coming from this simple static approach is clear. A decrease of the pore size implies the expected slowing-down of the average translocation time at fixed $f$. This effect, however, depends on the value of $f$, amounting to a modification of the overall shape of $\tau(f)$ mainly due to the presence of membrane-polymer interactions. Analogous modifications can therefore be expected in the case of the kinetic membrane, where the effective pore size $R_{\text{eff}}(t)$ is modulated in time, as we will see below.    
\subsection{Effect of nanopore deformation}
Pore {\em deformation} properties are obviously expected to play an important role on the translocation process. It has been demonstrated that, in the case of bilipidic membranes, the use of biological pores explicitly constraints the pore structure~\cite{Mereuta,Balijepalli}, which can therefore be considered completely immobile. Nano-pores carved in thick synthetic membranes are also considered as barely deformable~\cite{Dekker}. In the case of more recent experiments using graphene~\cite{Garaj, Merchant, Schneider} or other 2-dim crystals like MoS$_2$~\cite{Feng}, in contrast, deformation is expected to be important, especially for narrow pores. Membranes formed by 2-dim origami sheets are an additional emerging class of soft membranes where defomability upon polymer translocation may also play an important role~\cite{Bell, Hernandez, Wei}.

The effect of the dynamics of the membranes on the translocation process is highly non-trivial and, to the best of our knowledge, no general theoretical framework exists able to rationalize these conditions. Some numerical work has focused on the case where a time-dependent pore size is controlled in an oscillatory fashion~\cite{Cohen,Cohen2}, for instance. Here we consider the more realistic situation where an effective time-dependent pore radius is generated by the dynamics of the membrane itself. This is actually a quite complex dynamics, controlled by the interplay between the deformation of the pore due to the direct interaction with the translocating polymer beads, and the thermal agitation of the membrane components itself. We now try to disentangle these two effects. 
\begin{figure}[t]
\centering
\includegraphics[width=0.48\textwidth]{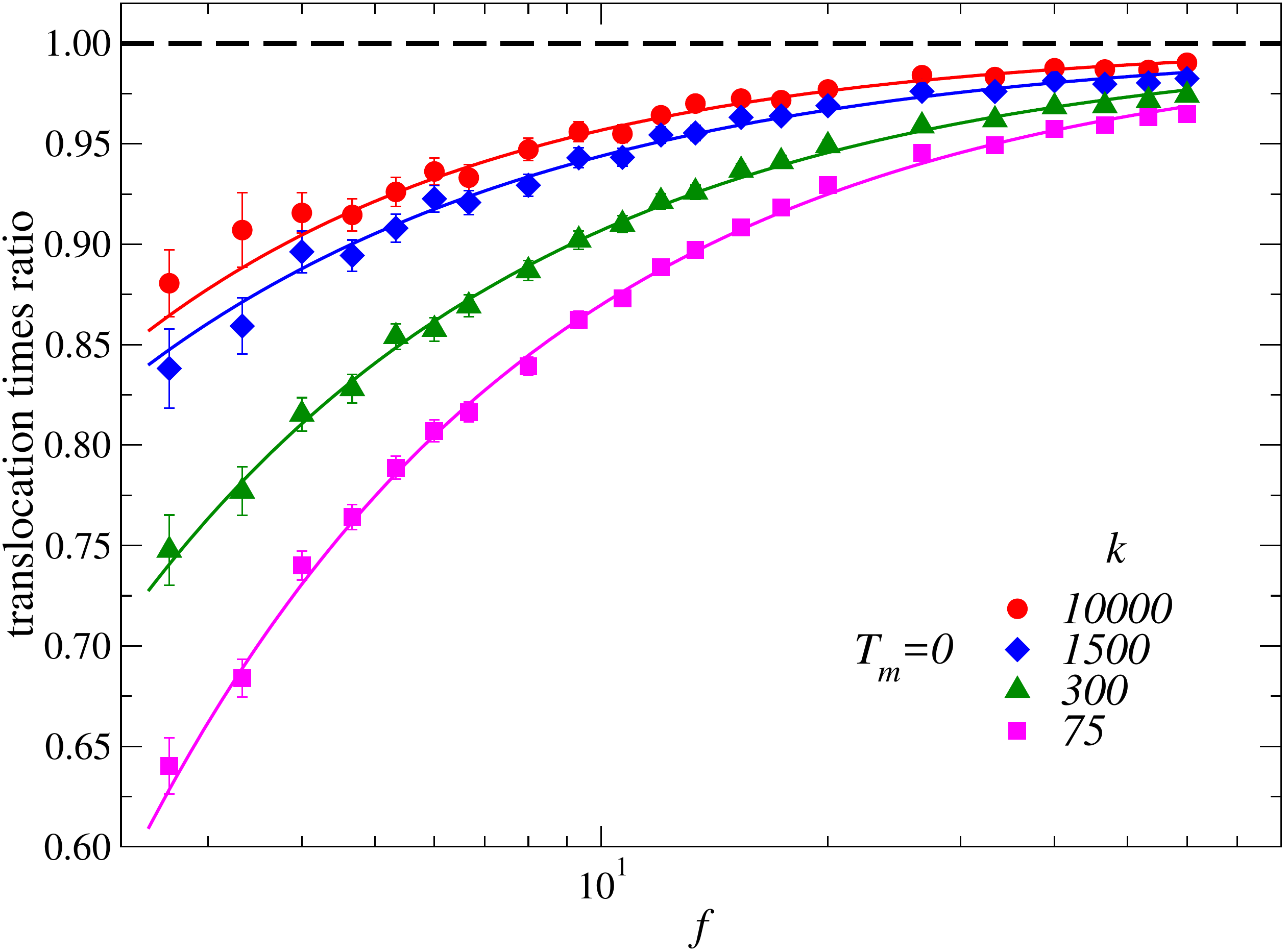}
\caption{
Renormalized average translocation time, $\tau^*=\tau(k,f)/\tau(\infty,f)$, at $T_m=0$. These data are the ratio of the average translocation time for different tethering (see main text) harmonic potential elastic constants $k$, to the fixed membrane (corresponding too $k = \infty$) reference value. The solid lines are guides for the eye, the dashed horizontal line is the reference value. The deformation of the nano-pore due to the direct interaction with the polymer beads reduces the average translocation time. The effect is stronger at lower values of $f$ and increases by decreasing $k$. A discussion of these data is included in the main text.
}
\label{fig:k}
\end{figure}

We first focus on this issue on the effect of the membrane deformability on the polymer translocation time $\tau$. As we expect stronger effects for smaller pores, we focus on the pore size $R_s$. The extent of the deformability is controlled by modulating the spring constant $k$ of the tethered membrane sites, at the vanishing membrane temperature $T_m=0$. We have considered values of $k$ in the range $75$ (high deformation) to $k = 10^4$ (low deformation). Note that the case of the immobile (fixed) membrane corresponds to $k = \infty$, which is our reference state. We plot our results in Fig.~\ref{fig:k}, where the values of the average translocation times are normalized to the corresponding value for the case of the fixed membrane, $\tau^*(k,f)=\tau(k,f)/\tau(\infty,f)$. Here the dashed line corresponds to the immobile limit and the solid lines are guides for the eye. (These latter are given by the sum of two power-laws, accounting for the low-$f$ and high-$f$ regimes respectively.) 

During translocation, we expect the membrane beads forming the pore to be displaced by the polymer beads occupying the pore at any time, amounting to an effective pore size which on average is slightly larger than the value in the immobile membrane. This is exactly what we have found by inspecting pore configurations during translocation. (We show a sub-set of our data in the Supplementary Fig. S5 online.) As a consequence, we find that the average translocation time is reduced compared to the immobile limit in all cases. In particular, $\tau$ are all very close at the higher values of $f$ where, again, translocation is a highly out-of-equilibrium process mainly controlled by $f$. The speed-up, in contrast, is higher at the lower pulling forces. Also, note that even a value of $k=10^4$ is not sufficient to recover the immobile limit $k=\infty$. Indeed, while the harmonic well constraining the pore beads very close to their equilibrium position is very steep, this still allows a substantial deformation of the pore which is characterized by an effective pore size which is larger than $R_s$. This amounts to a non-negligible speed-up of the translocation at low-$f$ with a maximum of $15\%$ at $f=2.5$. This effect is increasingly evident by decreasing $k$, therefore increasing the semi-plastic deformation of the pore. At the lowest value $k=75$ we find an overall speed up of about $35\%$ compared to the immobile case. 

The main message originating from these data is that, indeed, the nano-pore deformation has an important non-trivial effect, promoting translocation especially at the lowest values of the driving force. Unfortunately, even if the overall behavior of the data is similar at all values of $k$, we did not find any simple way to collapse all data on a single master curve. This means that, in the different conditions, the drive due to the pulling force and the extent of the allowed membrane pore deformation counter-balance without following any simple functional dependence on $f$ and $k$. Further quantitative insight on these data therefore implies the development of substantial theoretical work.  

We conclude this Section by noting that while the pore displacements induced by the low values of $k$ considered here are not consistent with the deformation properties of graphene, functionalization of the pore with increasingly flexible molecules (from alkanes to carbynes~\cite{LiuCarbyne}, for example) may add variable deformation properties to be tested in experiments. The large range of pore structures available for soft origami membranes~\cite{Hernandez2} also is a promising playground to explore in this direction.
\subsection{Influence of the membrane temperature}
In more realistic conditions, effects originating from thermal agitation of the membrane adds to those due to the deformability of the nano-pore investigated above. We obviously expect most part of the modifications to those data coming from the effect of temperature on the beads closer to the pore, and in direct interaction with the polymer. Non-negligible effects, however, are also certainly associated to membrane atoms further from the pore, exchanging momentum with the monomers before translocation. See additional data in Supplementary Figs. S4 and S5 online.)

\begin{figure}[t]
\centering
\includegraphics[width=0.48\textwidth]{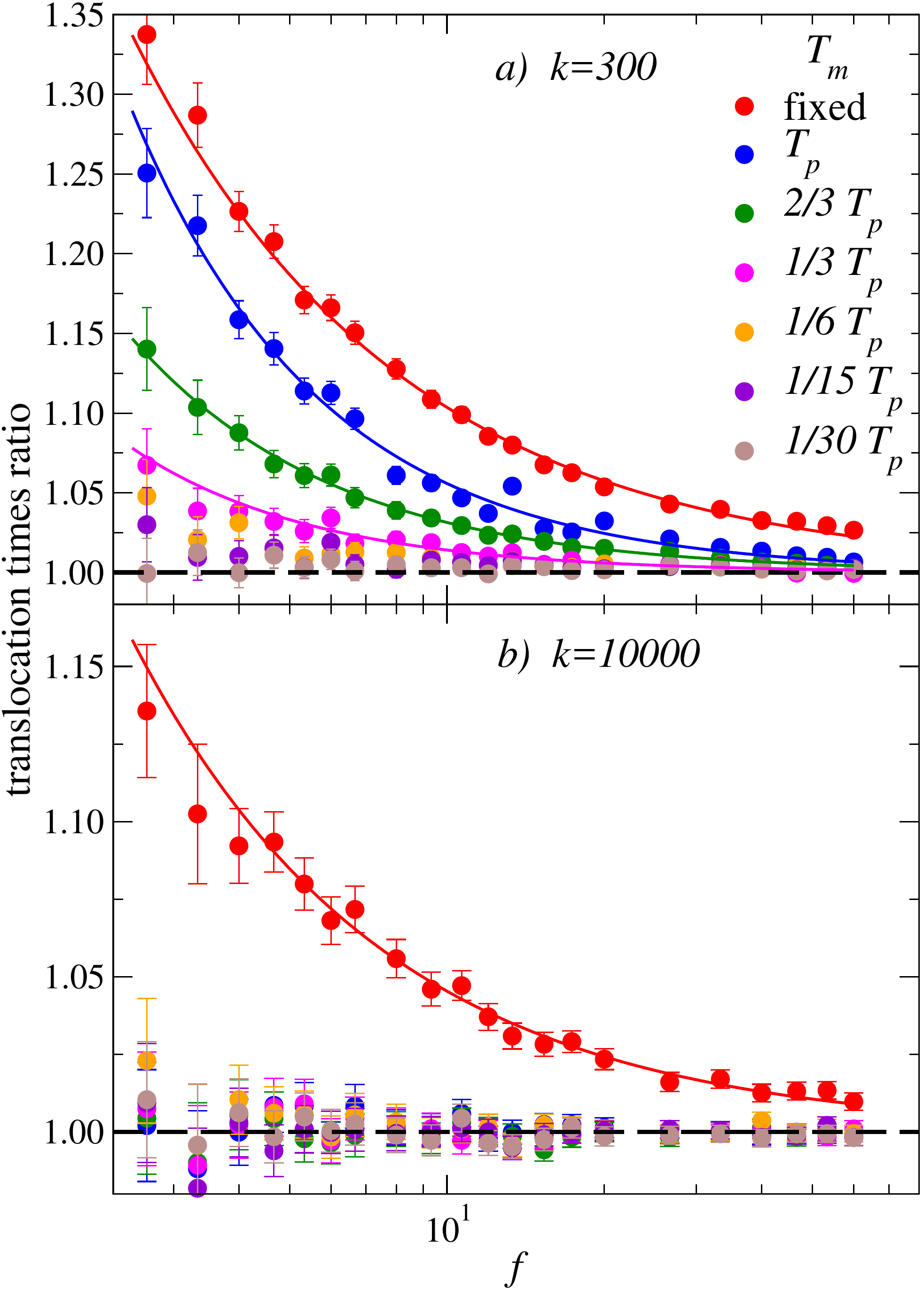}
\caption{
Renormalized average translocation time, $\tau^*=\tau(k,T_m)/\tau(k,T_m=0)$ at the indicated membrane temperatures. These data are the ratio of the average translocation time to the $T_m=0$ reference value. We show our data sets at two different values of the tethering harmonic potential elastic constants $k=300$ and $10^4$ in panel a) and b), respectively. The results for the fixed membrane are also shown in the two cases. Solid lines are guides for the eye, the dashed horizontal lines are the reference values. A discussion of these data is included in the main text.
}
\label{fig:T}
\end{figure}

We have therefore  analyzed the translocation process in the cases where the membrane is kept at a temperature $0< T_m\le T_p$. In Fig.~\ref{fig:T} we show our data at the indicated values of $T_m$, for $k=300$ (high deformability, top) and $k=10^4$ (low deformability, bottom). Here we show our data renormalized to the values at $T_m=0$ at each $f$ (see Fig.~\ref{fig:k}), $\tau^*(T_m,f)=\tau(T_m,f)/\tau(0,f)$. The dashed horizontal line therefore corresponds to this limit, while the solid lines are guides for the eye. (These latter are, again, given by the sum of two power-laws, accounting for the low-$f$ and high-$f$ regimes respectively.) We also plot the data corresponding to the immobile case, recalling that this does {\em not} correspond to the case $T_m=0$ for $k=10^4$.

In Fig.~\ref{fig:T}a) we show our data sets for $k=300$. For temperatures $T_m<T_p/3$ we recover the $T_m=0$ limit in the entire $f$-range, with barely detectable modifications at the lowest values of $f$. For  $T_m\ge T_p/3$, in contrast, data show a slowing down of the translocation process, which increases by decreasing $f$. This effect is temperature dependent, and is bounded from above by the immobile limit (red circles). Again, this finding can be rationalized by noticing that now the pore beads are also subjected to  thermal agitation, inducing an overall displacement $\propto T_m$, and amounting to a reduced effective pore size width. 

The data of Fig.~\ref{fig:T}b) corresponding to $k=10^4$ are also interesting. In this case, the mechanical constraints associated to the extremely steep tethering potentials are so strong that they completely overwhelm the effect associated to thermal fluctuations. As a consequence, the average translocation times stay very close to the $T_m=0$ case at all temperatures, well below the static limit fixed by the immobile membrane. The important difference observed between the data pertaining to the fixed membrane case and those corresponding to a very large value of $k$ and membrane $T_m$  down to very low values illustrate the effect of elastic vs inelastic collisions between membrane and polymer beads. Similarly to above, the picture is qualitatively very clear but a more extended theoretical understanding is needed for more quantitative insight. 
\subsection{Comparison with previous work}
It is instructive to compare our results on the effect of thermal vibrations with the recent studies of Refs.~\cite{Cohen,Cohen2}. In those works, a time-dependent nano-pore size has been induced by a controlled oscillating field at fixed frequencies. Interestingly, a translocation rate accelerated compared to that for the steady pore has been demonstrated, for a range of width oscillation frequencies. These findings have been interpreted in terms of a resonant activation of the translocation. Similar results have been reported in Refs.~\cite{Fiasconaro, Ikonen, Sarabadani}, where active polymer translocation was realized driven by an oscillating force~\cite{Fiasconaro, Ikonen, Sarabadani}, and reduced translocation times were also observed. 

These findings could seem in contrast with our results of Fig.~\ref{fig:T}, where a slowing-down (or invariance) of the translocation processes is observed in the entire range of membrane temperature and pulling force considered. This discrepancy, however, can be easily reconciled by noticing that in the case of Cohen et al.~\cite{Cohen,Cohen2} the nano-pore beads positions are driven cooperatively, originating a {\em coherent} breathing of the nano-pore which promotes translocation. This coherent breathing of the nanopore is also present in Sarabadani et al.~\cite{Sarabadani} approach. In our case, in contrast, thermal vibrations of the membrane beads lead to {\em incoherent} fluctuations of the pore width around the average pore size, with a negative impact on the translocation process. Furthermore, due to those {\em incoherent} thermal fluctuations, we may expect that the effective pore width $2 R_{\text{eff}}(t)$ is always reduced compared to the pore width $2 R_s$ with a zero temperature membrane. Interestingly, the translocation time $\tau(k,T_m)$ is however reduced compare to the fixed membrane results. This is mainly due to the deformability of the pore allowing a strong decrease of the translocation time that is not compensated by the thermal vibrations. 
\section{Discussion}
In this paper we have investigated by Molecular Dynamics simulation the driven translocation process of structured polymers through nano-pores carved in thin membranes. The coarse-grained model used for the polymer is meant to mimic a single-stranded DNA, while the membrane structure recalls graphene sheets used in recent experimental work. In contrast to previous numerical work, the beads forming the membrane are not immobile and are tethered to the equilibrium lattice positions by harmonic wells. They therefore undergo small displacements which are due both to direct interaction with the polymer beads, and to thermal fluctuations. The relative simplicity of the model allows for a plain statistical mechanics analysis of the translocation times probability distributions, which we have probed by generating a large ensemble of independent translocation processes driven by pulling forces of different strengths. 

We have first shown in the case of immobile membranes that reducing the size of the nano-pore slows down translocation at all values of the pulling force, as expected. More interestingly, however, we have demonstrated that this does not amount to a uniform shift of the time scale, but rather to an interesting cross-over between different behavior at high and low-$f$ for the smallest nano-pore. This is a convincing evidence that non-trivial effect can be expected in the kinetic case, where the effective size of the nano-pore fluctuates in time.

In order to disentangle the effect of pore deformation due to the passage of the polymer from the displacements driven by thermal fluctuations, we have first considered the case where the membrane temperature vanishes, while the tethering potential strength is tuned to mimic low and high deformability of the pore. We have demonstrated that this strongly impacts the average translocation times, with a substantial {\em speed-up} compared to the immobile case, whose rate increases with deformability. Interestingly, non-negligible thermal fluctuations of the membrane have an opposite effect, which overwhelms that associated to deformability of the pore and amounts to a {\em slowing-down} of the translocation process, with a rate which increases with temperature. These results have been next discussed in comparison with other recent numerical and experimental investigations.

This work constitutes, to the best of our knowledge, the first attempt to better understand the effect on the translocation process of the dynamics of the nano-pore due to thermal fluctuations and deformability properties of the membrane. Although the qualitative picture emerging from our data is clear, a deeper quantitative analysis is hampered by an insufficient available corpus of exact predictions, which would allow for a more rigorous description of the data of Figs.~\ref{fig:k} and~\ref{fig:T}. Substantial theoretical work is therefore needed in this direction. 

In conclusion, further extension of the present work include the more realistic case of a fully elastic sheet where the polymer translocation couples to the extended vibrational modes of the membrane, or functionalization of the nano-pore by molecules of tunable flexibility. This last set-up would allow a comparison of numerical data with recent related experimental work. By introducing selective interactions of these grafted molecules with the different polymer bases, it would also provide a way of directly simulating the process of sequence reading upon translocation. Recent results along those lines have shown the potential of base functionalization of nano-pores in graphene nanoribbons~\cite{Paulechka}.
\begin{acknowledgments}
We acknowledge the {\em Grand \'Equipement National de Calcul Intensif (GENCI)} for the allocation of computing resources INAC-SPRAM is part of the Arcane Labex program, funded by the French National Research Agency (ARCANE project n$^{\circ}$ ANR-12-LABX-003). 
\end{acknowledgments}
\bibliographystyle{apsrev4-1}
\bibliography{biblio}
\newpage


%


%

%
\end{document}